\documentstyle[epsf,12pt]{article}
\newcommand{\nobody}{\rule{0ex}{1ex}}

\newcommand{\req}[1]{(\ref{#1})}
\newcommand{\nn}{\noindent}
\newcommand{\nll}{\nonumber \\}

\newcommand{\bq}{\begin{equation}}
\newcommand{\eq}{\end{equation}}
\newcommand{\ba}{\begin{eqnarray}}
\newcommand{\ea}{\end{eqnarray}}
\begin{document}
\thispagestyle{empty}
\vspace{-1.4cm}
\begin{flushleft}
{LMU-05/97 \\}
 June 1997
\end{flushleft}
\vspace{2.0cm}
\begin{center}
{\LARGE  \bf
Radiation of two hard gluons in $e^+e^-\rightarrow q\bar q$
}\\
\vspace*{1.0cm}
\nn
{\large
M. Jack\footnote{email:\ \ jack@graviton.hep.physik.uni-muenchen.de}, 
A. Leike\footnote{email:\ \ leike@graviton.hep.physik.uni-muenchen.de}
        \footnote{Supported by the EC program CHRX-CT940579}
}
\\
\vspace*{0.5cm}

Sektion Physik der
Ludwig-Maximilians-Universit\"at,                 \\
Theresienstr. 37, D--80333 M\"unchen, Germany
\end{center}
\vspace*{1.5cm}
 
\centerline{\Large 
Abstract}
\vspace*{.3cm}
\nn
The process  $e^+e^-\rightarrow q\bar qgg$ where all final particles are
hard is treated in a semi-analytic approach. 
Cuts on the invariant masses of the quark and the gluon pair can be applied.
The total cross section is obtained by three numerical integrations.
 \vspace*{1.0cm}
\section{Introduction}
The study of $W$ pair production is an important
task of the present experiments at LEP \cite{l3mw}. 
The $W$ Bosons are only accessible through their decay products.
Therefore, the complete process $e^+e^- \rightarrow f_1\bar f_2f_3\bar f_4$
must be studied.
It can be calculated by Monte Carlo or semi-analytic techniques \cite{lep2proc}.
A Monte Carlo description is absolutely necessary to describe a real detector
and realistic experimental cuts.
A semi-analytic description allows only for simple cuts in the phase
space but it is very fast and still flexible enough for fits to data.
This is proved by the semi-analytic code {\tt GENTLE/4fan} \cite{gentle},
which is used by all four LEP collaborations to determine the $W$ mass
\cite{l3mw}. 

A jet due to a light quark cannot be distinguished experimentally from a
jet due to a gluon.
Hence the reaction 
\ba
\label{qqgg}
e^+(k_1)e^-(k_2) \rightarrow q(p_1)\bar q(p_2)g(p_3)g(p_4)
\ea
could give a signature similar to that of the reaction 
\ba
\label{qqqq}
e^+e^- \rightarrow q_1\bar q_2q_3\bar q_4.
\ea
Reaction \req{qqgg} is therefore an important incoherent background to
the process \req{qqqq}. 
The process \req{qqgg} can be measured in more detail when
the quark flavor $q$ can be tagged.
For heavy quarks, it is also a background to Higgs production \cite{higgs}.

The process \req{qqgg} was investigated many years ago
\cite{oldrefs} because a jet due to a light quark cannot be
distinguished from a jet due to the quark accompanied by soft and/or
collinear gluons. 
Monte Carlo studies of the process \req{qqgg} with four hard final
particles can be found in references \cite{mcqqgg}. 

The semi-analytic calculation of 4-fermion final states
neglecting quark masses leads to very compact formulae 
for double and triple differential cross sections \cite{ringberg}.
Only one (two) numerical integrations are necessary to evaluate simple
distributions (cross sections).
This makes the semi-analytic codes fast compared to Monte Carlo programs.

The aim of this paper is the derivation of semi-analytic formulae
describing the process \req{qqgg}, which will be added to the code
{\tt GENTLE/4fan} \cite{gentle} in order to enhance the flexibility 
of this code.
In the calculation of reaction \req{qqgg}, we must apply
the same cuts as to the reaction \req{qqqq}.
In particular, the invariant energies of the gluon and quark pairs 
in \req{qqgg} are bounded from below.
This is necessary to keep the cross section out of the range of
hadronic resonances where perturbation theory breaks down.
Such cuts also ensure that the gluons cannot become soft.
Where all four final particles are hard, we can limit
ourselves to a tree-level calculation. 

The numerical largest corrections due to the radiation of photons
from the initial state can be taken into account by a convolution ,
\bq
\sigma^{ISR}(s)=\int_{s'^-}^s \frac{d{s'}}{s}\sigma^0(s')\rho(s'/s).
\eq
See reference \cite{gentle} for further details and references.
\section{Calculation}
We parameterize the phase space by 6 angles and 2 invariants,
%
\ba
\label{domega}
d\Gamma
&=& \prod_{i=1}^4\frac{d^3p_i}{2p_{i}^0}
 \times  \delta^4(k_1+k_2-\sum_{i=1}^4 p_i)
\nonumber
\\
&=&\frac{\sqrt{\lambda(s,s_q,s_g)}}{8s}
\frac{\sqrt{\lambda(s_q,m_1^2,m_2^2)}}{8s_q}
\frac{\sqrt{\lambda(s_g,m_3^2,m_4^2)}}{8s_g}
d s_q d s_g d \Omega_0 d \Omega_q d \Omega_g,
\ea
with 
\ba
\label{lambda}
\lambda(a,b,c) &=& a^2+b^2+c^2-2ab-2ac-2bc,\ \ \ 
\lambda \equiv \lambda(s,s_q,s_g),\nll
s&=&(k_1+k_2)^2,\ \ \ s_q=(p_1+p_2)^2\mbox{\ \ and\ \ }s_g=(p_3+p_4)^2,\nll
p_i^2&=&m_i^2,\ \ \ m_1=m_2=m_q,\ \ \ m_3=m_4=m_g=0.
\ea
The solid angles $\Omega_q$ and $\Omega_g$ are defined in the rest frames of
the quark and gluon pairs. 
$\Omega_0$ is the solid angle of $\vec p_1+\vec p_2$ in the laboratory frame.
We have $d\Omega_i = d\cos\theta_id\phi_i$ for $i=0,q,g$.
The kinematical ranges of the integration variables are
\ba
(m_1+m_2)^2 \le s_q \le (\sqrt{s}-m_3-m_4)^2&,&\ \ \ 
(m_3+m_4)^2 \le s_g    \le (\sqrt{s}-\sqrt{s_q})^2 ,
\nll
-1 \le \cos\theta_i \le 1 &,&\ \ \ 
0 \le \phi_i \le 2\pi.
\ea
%

The process \req{qqgg} is described by 16 Feynman diagrams, 8 with virtual
photon exchange and 8 with virtual $Z$-Boson exchange.
6 of the 8 diagrams are obtained by attaching the two gluons in all possible
combinations to the quark pair, the other two diagrams contain the 
triple-gluon vertex. The two types of diagrams are presented below in Figure~1.

%
%
\begin{figure}[htb]
\begin{minipage}[bht]{7.9cm}
{\begin{center}
  \vspace{7.0cm}
  \hspace{-5.0cm}
  \mbox{
  \epsfysize=14cm
  \epsffile[0 0 500 500]{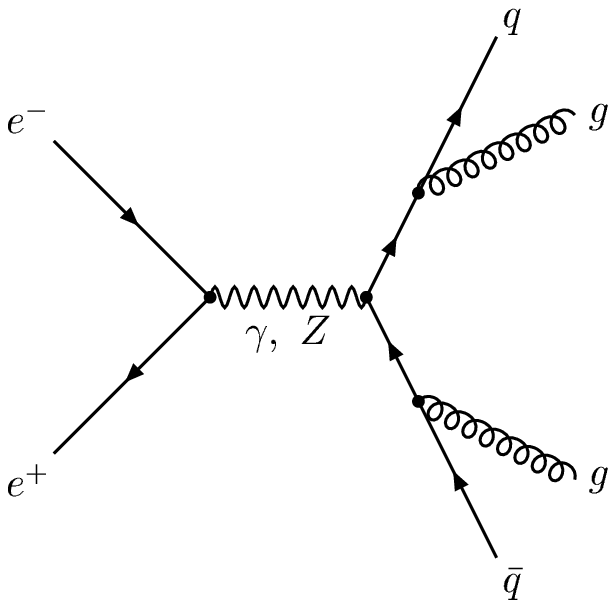}
  }
\end{center}}
\end{minipage}
\begin{minipage}[bht]{7.9cm}
{\begin{center}
  \vspace{7.0cm}
  \hspace{-2.0cm}
  \mbox{
  \epsfysize=14cm
  \epsffile[0 0 500 500]{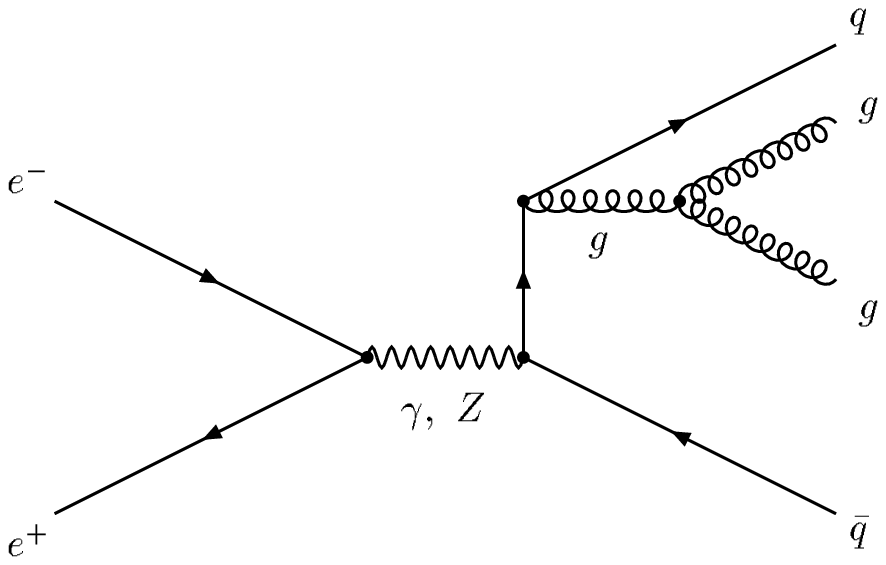}
  }
\end{center}} 
\end{minipage}
\vspace{-15.0cm}
{\small\it\nobody\hspace{3cm}
{\bf Figure 1:}
Radiation of the 2 gluons by the quarks
and triple-gluon vertex}
\end{figure}

We obtain distributions and cross sections by integrating over the
matrix element squared, which is obtained by traditional trace techniques.
Our semi-analytic approach does not allow for angular cuts to separate
gluons from quarks. 
Therefore, we keep a non--zero quark mass to avoid collinear
singularities. 
The calculation is done with the help of the symbolic manipulation program
{\tt FORM} \cite{form}. 

Without transversal beam polarization, the integration over the angle
$\phi_0$ is trivial giving $2\pi$. While the first 2 integrations over
$\phi_q$ and $\phi_g$ as azimuthal angles of the  
$q\bar{q}$- and $gg$-subsystems in the chosen parameterization can be 
done relatively easily and the integration over the solid angle 
$\cos\theta_0$ is trivial because only terms proportional to $1,\ \cos\theta_0$
and  $\cos^2\theta_0$ arise, the last 2 angular integrations prove to
be more complicated \cite{diploma}. 
This is due to logarithms with quadratic or biquadratic polynomials 
in $\cos\theta_{q,g}$ in their arguments after one integration. 
In order to avoid singularities for the last 2 angles $\cos\theta_{q,g}$
-- these reflect the collinear singularities of the original amplitude
for $m_q = 0$ -- the quark mass $m_q$ cannot be neglected. So from
these last 2 integrations one can be done analytically leading to 
dilogarithms of the type $Li_2(s_q,s_g;s,m_q^2)$\ and the above 
mentioned logarithmic integrals as dominant contributions, while the 
other integration is treated numerically. An analytical integration
over $\cos\theta_g$ in the massless $gg$-subsystem with a numerical 
integration over $\cos\theta_q$ afterwards leads to shorter
expressions of the integrand than vice versa and is favored.

The total cross section may then be written in the form
\ba
\label{tsig}
\sigma(s) = 
\int_{{\bar s}_q}^s ds_q
\int_{{\bar s}_g}^{(\sqrt{s}-\sqrt{s_q})^2} ds_g
\int_{-1}^1d\cos\theta_q\,
 \frac{\sqrt{\lambda}}{\pi s^2}\,
C_{qqgg}(e;q;s){\cal G}_{qqgg}(s_q,s_g,\cos\theta_q;s,m_q^2)\quad , 
\ea
with numerical integrations over $\cos\theta_q,\ s_q$ and $s_g$.
In the integration (\ref{tsig}), we allow for cuts $\bar s_q$ 
and $\bar s_g$ on the invariant masses.
After neglecting quark masses, the function $C_{qqgg}$ containing couplings
and gauge boson propagators, is
\ba
\label{cqqgg}
C_{qqgg}(e;q;s) &=& \frac{2}{(6\pi^2)^2}\alpha_s^2(4\pi)^2 \;
  \Re e   \sum_{V_i,V_j=\gamma,Z}
\frac{1}{D_{V_i}(s)}\frac{1}{D^*_{V_j}(s)}
\\ & &
\times~
\left[ L_e(V_i)L_e(V_j)+R_e(V_i)R_e(V_j)\right]
\left[ L_q(V_i)L_q(V_j)+R_q(V_i)R_q(V_j)\right],\nll
D_V(s)&=&s-M_V^2+iM_V\,\Gamma_V.\nonumber
\ea
$L_f(V)$ and $R_f(V)$ are the left- and right-handed couplings of the
fermion $f$ to the vector Boson $V$ with the mass $M_V$ and the width
$\Gamma_V$. The preceding factor is kept for convenience with reference
\cite{gentle2}. 

The analytic expression for the function ${\cal G}_{qqgg}(s_q,s_g,
\cos\theta_q;s,m_q^2)$ is much longer than those obtained 
for the four-fermion final states investigated in \cite{gentle2}.
This is partially due to the richer topology of the Feynman diagrams,
partially due to the non-zero quark mass appearing as an additional
dimensional parameter and partially due to the fact that the
integration over $\theta_q$ cannot be done analytically.
As was recognized in \cite{ringberg}, the analytic expressions
always tend to be longer before the integration over
the last angle is done.

Several checks are applied to ensure the validity of our calculation.
The matrix element must satisfy Bose symmetry.
In addition, it must vanish when the polarization vector of anyone of the
gluons is substituted by the momentum vector of this gluon.
All ``artificial'' poles appearing from partial fraction decomposition must
be compensated by corresponding terms in the nominator.
In particular, the result must be finite in the limit $\lambda\rightarrow 0$
and $\cos\theta_i\rightarrow\pm 1$.
Taylor expansions are necessary in different regions of the phase space to
obtain results, which are numerically stable. 
All analytical integrals entering our calculation are checked numerically.
In addition, our analytic result is checked numerically against our 
amplitude squared with a 7-fold integration by the integration routine
{\tt VEGAS} \cite{vegas}. Finally, we compared our result numerically
with {\tt COMPHEP} \cite{comphep} for different c.m. energies, quark 
masses and cuts. 
%
\section{Results}
%
\begin{figure}[tbh]
\ \vspace{1cm}\\
\begin{minipage}[t]{7.8cm} {
\begin{center}
\hspace{-1.7cm}
\mbox{
\epsfysize=7.0cm
\epsffile[0 0 500 500]{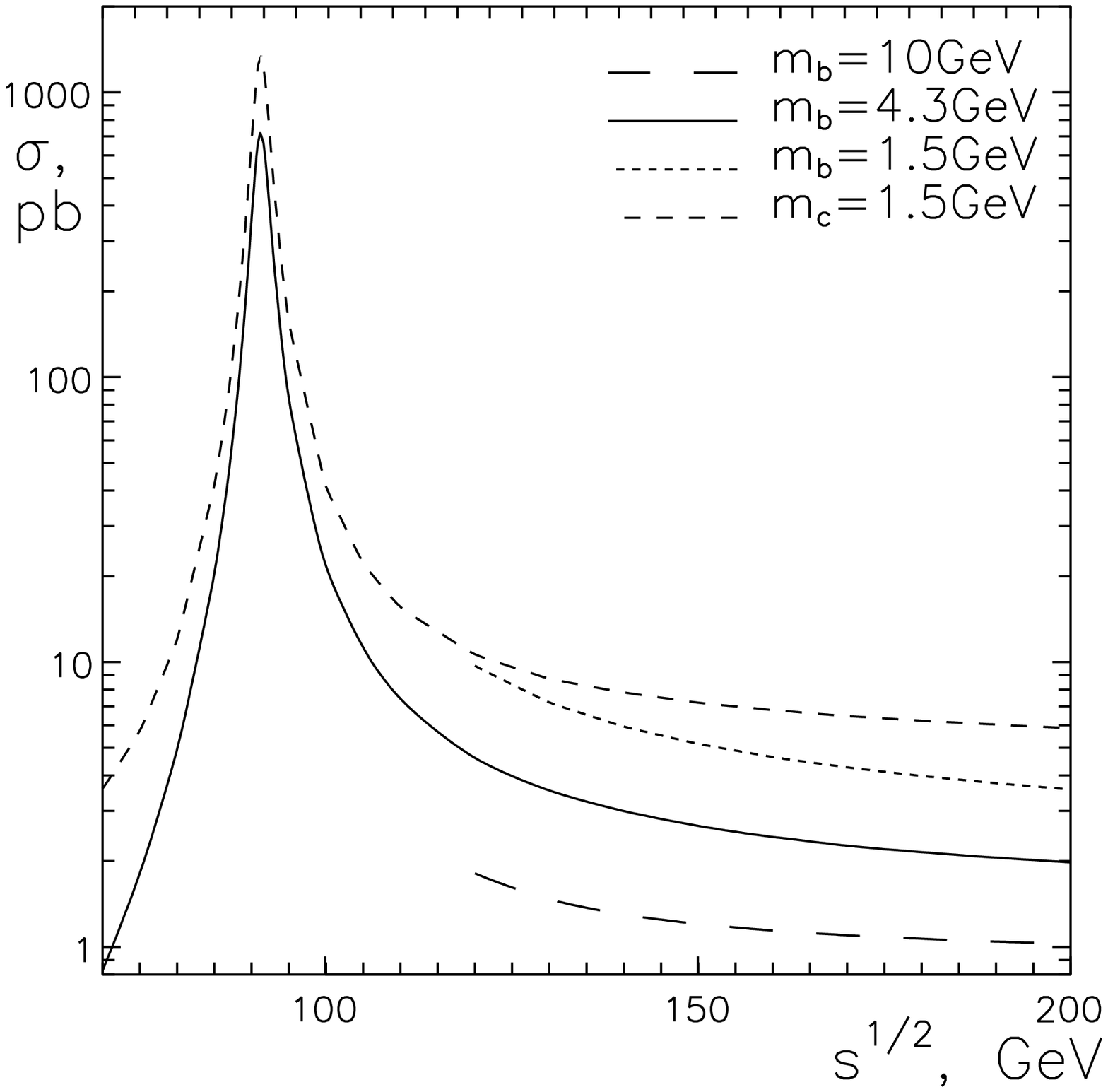}
}
\end{center}
\vspace*{-0.5cm}
\noindent
{\small\it
{\bf Figure 2:}
The total cross section \req{qqgg} as a function of the c.m. energy for
$m_b=4.3\,GeV$ and $m_c=1.5\,GeV$.
For $\sqrt{s}>120\,GeV$, the cross section is also shown for $m_b=1.5\,GeV$
and $m_b=10\,GeV$.
The cuts $\sqrt{\bar s_q}=\sqrt{\bar s_g}=20\,GeV$ are applied.
}
}\end{minipage}
\hspace*{0.5cm}
\begin{minipage}[t]{7.8cm} {
\begin{center}
\hspace{-1.7cm}
\mbox{
\epsfysize=7.0cm
\epsffile[0 0 500 500]{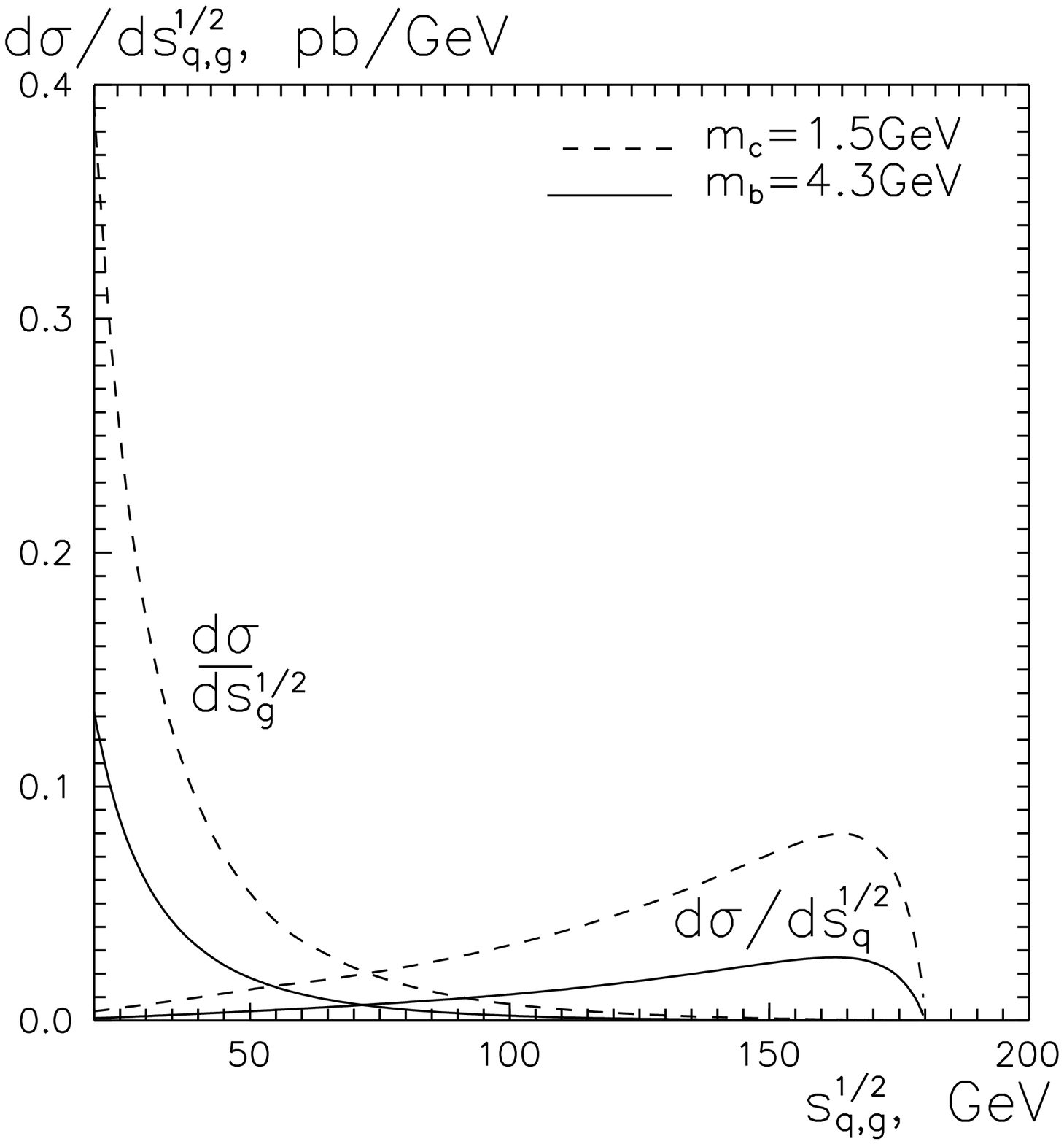}
}
\end{center}
\vspace*{-0.5cm}
\noindent
{\small\it
{\bf Figure 3:}\hfill
The differential distributions\\ 
$d\sigma/d\sqrt{s_q}$ and $d\sigma/d\sqrt{s_g}$ 
for $\sqrt{s}=200\,GeV,\ m_c=1.5\,GeV$ and $m_b=4.3\,GeV$. The cuts
$\sqrt{\bar s_q}=\sqrt{\bar s_g}=20\,GeV$ are applied.  
}
}\end{minipage}
\end{figure}

The numerical input for our figures is 
$\alpha_s=0.121,\ \alpha=1/128,\ \sin^2\theta_W=0.232,\ M_Z = 91.186\,GeV,$
and $\Gamma_Z=2.495\,GeV$.

The total cross section for $q=b,c$ is shown in Figure~2.
To illustrate the mass dependence, the cross section is given for different
$b$-quark masses beyond the $Z$ peak.
The considered process has a large cross section.
For light quark masses and large energies $\sqrt{s}\ge 200\,GeV$, it
becomes comparable to the cross section of $e^+e^-\rightarrow q\bar q$.
This is due to the radiation of two {\it collinear} gluons.
For $\sqrt{s}=200\,GeV$ and $m_q=1.5\,GeV$, every collinear
gluon generates a logarithm $\ln\frac{s}{m_q^2}\approx 9.8$, which
numerically compensates the coupling constant $\alpha_s$. 
Other contributions to the cross section are of comparable importance
for lower energies or larger quark masses.
The logarithmic mass terms are the reason for the increase of the total
cross section with smaller quark masses.

Figure~3 shows the differential distributions of the invariant energies of
the quark and gluon pairs in the final state.
The distributions are zero for $\sqrt{s_i}<20\,GeV$ and $\sqrt{s_i}>180\,GeV$
due to our kinematical cuts.
The distributions for $\frac{d{\sigma}}{d{\sqrt{s_g}}}$ peak for small $s_g$
reflecting the infrared peak due to the radiation of soft gluons. 
The distributions for $\frac{d{\sigma}}{d{\sqrt{s_q}}}$ are largest for
values of $s_q$ near the upper limit $(\sqrt{s}-\sqrt{\bar{s}_g})^2$
again due to the radiation of soft and collinear gluons.

{\em To summarize},
we performed a semi-analytic calculation of the process
$e^+e^-\rightarrow q\bar qgg$ where all final particles are assumed to be hard.
The process has a large cross section.
It is a large incoherent background to the four-fermion processes currently
measured at LEP.
\section*{Acknowledgement}
We would like to thank D. Bardin for discussions and T. Riemann for stimulating
this work, discussions and the careful reading of the manuscript.
%

\end{document}